# An Efficient Flow-based Multi-level Hybrid Intrusion Detection System for Software-Defined Networks

**Majd Latah** [1,*], **Levent Toker** [2]

[1] Department of Computer Science, Ozyegin University, Istanbul, Turkey; majd.latah@ozu.edu.tr
[2] Department of Computer Engineering, Ege University, Izmir, Turkey; levent.toker@ege.edu.tr
* Correspondence: majd.latah@ozu.edu.tr; Tel.: +90-543-7720219

**Abstract:** Software-Defined Networking (SDN) is a novel networking paradigm that provides enhanced programming abilities, which can be used to solve traditional security challenges on the basis of more efficient approaches. The most important element in the SDN paradigm is the controller, which is responsible for managing the flows of each correspondence forwarding element (switch or router). Flow statistics provided by the controller are considered to be useful information that can be used to develop a network-based intrusion detection system. Therefore, in this paper, we propose a 5-level hybrid classification system based on flow statistics in order to attain an improvement in the overall accuracy of the system. For the first level, we employ the k-Nearest Neighbor approach (kNN); for the second level, we use the Extreme Learning Machine (ELM); and for the remaining levels, we utilize the Hierarchical Extreme Learning Machine (H-ELM) approach. In comparison with conventional supervised machine learning algorithms based on the NSL-KDD benchmark dataset, the experimental study showed that our system achieves the highest level of accuracy (84.29%). Therefore, our approach presents an efficient approach for intrusion detection in SDNs.

## 1. Introduction

Software-defined networking (SDN) is the fruit of earlier proposals that mainly concerned both programmable networks and control-data planes's separation [1]. In the SDN paradigm, the controller, which represents a logically centralized controlling point, successfully manages and gathers flow-based statistics via southbound interface such as OpenFlow protocol [2], which is maintained by Open Networking Foundation (ONF), a non-profit industry consortium that offers support and ensures various improvements in the SDN field [2]. In this paper, we aim to design an efficient Intrusion Detection System (IDS) in the SDN paradigm. Our goal is to correctly classify well-known attacks alongside being able to detect new attacks on the basis of flow statistics provided by the controller. Flow-based intrusion detection approaches [3–5] depend only on the inspection of the packet header; therefore, they are considered to be computationally efficient in comparison with packet-based systems that require the analysis of the packet's payload [6]. In addition, packet-based systems cannot be used when network traffic is encrypted [7]. Conversely, flow-based approaches cannot be utilized when the attack is embedded into the packet's payload [8]. Therefore, packet and flow-based approaches can be employed together in order to achieve the highest level of protection [9–10]. Moreover, multi-layer approaches [11–15, 17, 18, 20–27] serve as a potential solution to enhance the overall accuracy of intrusion detection systems. In this paper, we utilize a multi-layer approach based on the K-Nearest Neighbor (kNN), the Extreme Learning Machine (ELM), and the Hierarchical Extreme Learning Machine (H-ELM) classification methods. Our contribution is four-fold and can be summarized in the following manner:

- Designing a multi-level hybrid approach that permits intrusion detection only on the basis of 6-flow features that can be easily obtained by a typical SDN controller.
- Employing machine learning-based classification algorithms that reduce the time required for the testing stage, such as ELM and H-ELM.
- Testing the system through the use of a standard dataset (NSL-KDD) that includes a set of new attacks that did not appear in the training dataset, in order to validate the effectiveness of our proposed system.
- Accuracy improvements, from 75.75% to 84.29%, in comparison with well-known supervised machine learning approaches that employ the same 6-flow-based features and dataset.

The rest of this work is organized in the following manner. Related work of multi-level IDS is presented in Section 2. The backgrounds of kNN, ELM, and H-ELM are introduced in Section 3. The proposed 5-level hybrid classification system is explained in detail in Section 4. The dataset used in training and testing stages is reviewed in Section 5. The experimental results are discussed in Section 6. Evaluation metrics are introduced in Section 7. Finally, the paper is concluded in Section 8.

## 2. Related Work

In [11], intrusions were classified through the use of a 3-level classification model based on C4.5 algorithm. The first stage classifies the test data into their corresponding attack group (DOS, PROBE, Others or Normal). The second stage classifies "Others" into U2R and R2L groups. The third stage categorizes the attacks into a one specific attack type (e.g. back, land, etc.). The model achieved an 88.3% accuracy level with considerably high false alarm rate that reached 11.158%. Detection rate for known attacks reached 84.312%. However, it reached 42.002% for unknown attacks, which is considered to be the main limitation for this model.

MLIDS [12], on the other hand, achieved a high detection rate (99.986%) and minimized the false alarm rate to almost zero through the analysis of network traffic on the basis of three levels of granularities (rule-based flow analysis, protocol behavior analysis, and payload behavior analysis) and the utilization of an efficient fusion decision algorithm on the basis of least squares technique for each connection key and time-window. Since it includes a sliding time-window and payload analysis, it is likely to introduce a burden on the network, and therefore, it is used for the purpose of analysis.

Abuadlla et al. [13] proposed a two-stage intrusion detection and classification method by taking advantage of two separate neural networks for each task. The first stage detects traffic anomalies whereas the next stage classifies the attacks as they occur. Experimental studies were conducted on the DARPA 1998 intrusion detection dataset. In the first stage, the detection rate reached 94.2% with a false alarm of 3.4%. Conversely, the second stage displayed a significantly high detection rate (99.42%) with a false alarm of 0.32%. The detection rate for known-attacks reached 99.97%, whereas it reached 78% for unknown attacks. It is worth noting that Multi-layer Perceptron (MLP) with Levenberg-Marquardt takes less training time in comparison with Resilient Backpropagation alongside having low memory consumption in comparison with Radial Basis Function. Despite mainly using flow-based features that can easily be collected from traditional network routers through the utilization of standard protocols (NetFlow, Jflow, IPFIX) [13], this approach uses in its first stage (SYN - SYN/ACK) features, where it should inspect the header of each packet in order to extract related TCP flags, and therefore, it represents an expensive feature collection that, in turn, could be an additional burden on the network.

Authors of [14] proposed a two-stage intrusion detection system with the employment of the Support Vector Machine (SVM) as an anomaly at the first stage and the Artificial Neural Network (ANN) as a misuse detector at the next stage. Experimental studies were conducted on the NSL-KDD intrusion detection dataset. The first stage classifies the network traffic into normal and attack groups. The next stage classifies the attack traffic into four attack groups. In the first stage, the detection rate reached 99.97% with a false alarm of 0.19%. Conversely, the second stage showed that the detection rate had reached 99.9% with a lower false alarm of 0.1%. Experimental results in this study reveal that misuse detection techniques are useful for the detection of known attacks with a low false positive rate.

In another study, Amoli and Hamalainen [15] proposed a real-time multi-stage intrusion detection system in order to enhance the detection rate of unknown attacks. The first stage detects potential attacks by monitoring a set of traffic features that include the size and number of bytes, packets, and network flows. The second stage, alternatively, finds the similarities with previous communications established by the intruders. If the number of any of the previously mentioned features crosses the threshold, a specific time slot will be flagged as anomaly. In order to increase the detection rate, the system monitors the rate of time difference between each packet (TDP) and network flows (TDF). Sub-space clustering is used to unsupervisedly cluster the previously flagged traffic, which allows the differentiation of normal traffic through the detection of the outliers with lower processing time and complexity in comparison with multidimensional clustering algorithms [15]. After obtaining several two-dimensional clusters, DBSCAN [16] is used for to create the proper clusters and to identify the outliers. The second stage is designed to discover potential similarities between intruders, which would be able to find the Bost-Master in case of Botnet attacks. To this end, past hour records of related network traffic should be stored. It is obvious that the system includes the use of multiple time windows and therefore, could negatively affect the performance of the controller if it is considered to be used in the SDN paradigm.

Aziz et al. [17] proposed a three-stage hybrid intelligent approach. The first stage employs the principal component analysis (PCA) method for feature selection, selecting 22 out of 41 features from the NSL-KDD benchmark dataset. The second stage applies genetic algorithm in order to generate anomaly detectors that are used to identify normal and anomalous behaviors. The last layer utilizes different classifiers (Naive bayes, MLP neural network, and decision trees) in order to increase the detection accuracy. The experimental studies showed that naive bayes classifier has better accuracy in detecting U2R and R2L attacks. Conversely, the J48 decision tree classifier achieves a good accuracy (82%) in detecting DoS attacks and 65.4% in detecting probe attacks.

Cordella et al. [18] proposed a serial multi-stage intrusion detection system on the basis of the Learning Vector Quantization (LVQ) classifiers, where the authors use the reliability concept [19] in order to assist the decision making at each stage. In case of low reliability, the system will not raise an alert which, in turn, will aid the reduction of the false alarm rate. A connection will be declared as normal traffic only if all stages do not detect any attack. The first stage distinguishes between normal traffic and the attacks belong to DoS or Probe classes. The second stage discriminates between normal traffic and R2L attacks. And finally, the last stage classifies the network traffic into normal and U2R attack classes. The overall error and missed detection reached 0.68% and 0.67% respectively.

Gogoi et al. [20] proposed a three-level intrusion detection system that utilized a combination of supervised, unsupervised, and outlier-based techniques for the enhancement of detection accuracy. The first level classifies test data into three classes: DoS, Probe, and Rest (unclassified), on the basis of an improved version of the CatSub algorithm [21]. The next level categorizes the Rest into Normal and Rest classes depending upon the k-point algorithm [22]. The last level splits the Rest into U2R and R2L classes. The data of the U2R class were extracted from the Rest class through the employment of an outlier-based classifier [23]. The remaining records in the Rest class were classified as R2L. The efficiency of this model was demonstrated by experimental studies conducted on four datasets, namely TUIDS (packet and flow level), DDoS, KDD Cup 99, and NSL-KDD. It is worth observing that the system was able to achieve high accuracy reaching 99.3%, 99.2%, 98.901%, 97.568%, and 98.394% on the TUIDS-packet, TUIDS-flow, DDoS, KDD Cup 99, and NSL-KDD datasets respectively. Alternatively, false positive rate on the previously mentioned datasets reached 0.0108%, 1.36%, 0.3505%, 9.933%, and 1.585% respectively.

Reddy and Achary [24] proposed a two-stage system on the basis of a combination of rule-based classifier and K-means clustering. Experimental results based on the KDD Cup 99 dataset displayed a good accuracy reaching 83.96%. Lee et al. [25] proposed a multi-stage agent-based intrusion detection system that utilizes the Hidden Markov Model (HMM). Based on experimental studies conducted on the DARPA 2000 intrusion detection dataset, detection rates were higher than 90% and duration times were below 30 ms. Rajeswari and Kannan [26] proposed a multi-stage system on the basis of the enhanced C4.5 algorithm. Detection rate based on the KDD Cup 99 dataset reached 62.33%, with a 9.1% false- alarm rate.

Araki et al. [27] employed a multi-stage One-Class Support Vector Machine (OC-SVM) system to detect sophisticated unknown attacks, wherein the system makes use of three sets of traffic, two sets retrieved from a dataset and one extracted from the real network. At the first stage, the OC-SVM learns from older archive sets and later analyzes a newer set and one from the real network. At the second stage, the OC-SVM learns the outliers from the newer set and utilizes them in order to analyze the outliers of the real network. The detection rate and the false positive rate reached 80.00% and 20.94% respectively.

Al-Yaseen el al. [29] proposed a multi-level hybrid system on the basis of SVM and ELM. Based on experimental results conducted on the KDD Cup 99 dataset, the system showed high accuracy (95.75%) and low false alarm rate (1.87%). Casas et al. [29] proposed a three-step clustering technique on the basis of sub-space density clustering in order to detect outlying flows where they use the following features: number of source/destination IP addresses and ports, ratio of number of sources to number of destinations, packet rate, fraction of ICMP and SYN packets, and average packet size. The first step detects anomalous time slots in which the analysis be conducted. The next step identifies and ranks the outlying flows according to the degree of abnormality on the basis of a combination of sub-space clustering [30], DBSCAN [16], and evidence accumulation clustering [31]. Thereafter, the top-ranked flows are considered as anomalies on the basis of a thresholding approach. The approach proposed by Casas et al. [29] displayed a better performance in comparison with other unsupervised anomaly detection methods such as DBSCAN, k-means, and PCA. However, since it extracts some TCP flags such as SYN, it may not be appropriate for the SDN case due to the fact that dealing with flows is more fast and efficient than handling each packet separately.

Previously discussed studies depend mainly on the use of the whole dataset that probably contains different features (basic features, time-related features, connection-related features, content-related features, host-related features, and login attempts-related features). Using all these features, however, may not be the best choice for designing a network-based IDS in the SDN paradigm. Recently, a flow-based deep learning approach [32] has been proposed for the purpose of intrusion detection in SDNs, where the system achieved a good accuracy reaching 75.75% only on the basis of 6-flow features. The same features also are used in our proposed system. The usage of flow-based features is considered straightforward and very effective in the SDN architecture where the SDN controller is already able to obtain such features without any need for a preprocessing or flow aggregation step. In this study, we show that our proposed system, which employs a flow-based multi-level approach, can achieve better results as compared by [32] and other conventional supervised learning approaches.

**3. Theoretical Background**

In this section, we introduce the theoretical background of the algorithms used in our proposed system. Therefore, we describe in details the following algorithms: kNN, ELM, and H-ELM.

*3.1. K-Nearest Neighbor algorithm (kNN)*

K-Nearest Neighbor algorithm (kNN) classifies the new instances that exist in a given dataset according to their closest training instances in the feature space [33]. kNN is a straightforward algorithm and displays a good robustness against noisy training data or a large dataset [34]. It is worth mentioning that kNN and its variants [35–39] are used widely in malware detection [40, 41], intrusion detection [42–45], and spam detection [46]. Typically, the algorithm finds the $k$ closest instances based on a calculation of the distance between the new instances and all training instances. For instance, let X and Y be two feature vectors with dimension $n$. The Euclidian distance between these two feature vectors is defined below

$$d_{XY} = \max(|X_i - Y_i|), \; i \in n \tag{1}$$

The new instance is classified based on the majority vote of its $k$ nearest instances. Therefore, it will be assigned to the class whose labels are the most frequent. Other distance metrics such as Manhattan, Mahalanobis, Chebyshev, Minkowski, and Hamming can be utilized as well. The performance of kNN is affected by the distance metric used by the algorithm alongside the choice of the optimal value of the parameter $k$ [47]. A small k increases the impact of individual cases. A large $k$, however, increases the robustness of the noise provided in the dataset [48]. The value of $k$ is commonly determined through the employment of cross-validation or adaptive approaches [49]. In this study, we use the standard version of this algorithm by using the Euclidian distance as the distance metric and applying cross-validation in order to determine the optimal value of parameter $k$. kNN algorithm is used in the first layer of our proposed model in order model for the detection of DoS attacks.

*3.2. Extreme Learning Machine (ELM)*

The Extreme Learning Machine (ELM) [50] represents a new learning paradigm built on basis of the concept of single-hidden layer-feed-forward neural network. ELM requires short training and testing time in comparison with traditional feed-forward networks owing to the fact that the parameters of its hidden layer are randomly chosen, thereby eliminating the need for the training stage [51-53]. ELM achieved high accuracy for multi-level IDS as compared with SVM [52]. Kernel-based ELM outperformed neural networks approaches for detecting Probe attacks in a multi-level IDS model [53]. For a given data set $Z$ $\{(x_1, t_1), (x_2, t_2), \ldots, (x_1, t_1) : i = 1, \ldots, N\}$, where $x_i = [x_{i1}, x_{i2}, \ldots, x_{in}]^T \in R^n$ and $x_i = [x_{i1}, x_{i2}, \ldots, x_{im}]^T \in R^m$, which is the label of instances, a single-hidden layer feed-forward neural network (SLFN) with $n$ inputs, $m$ outputs, $k$ hidden neurons, and activation function $g(x)$ can be written as

$$\sum_{i=1}^{k} \beta_i g(w_i^T x_j + b_i) = t, \quad j = 1,..,N \tag{2}$$

where $w_i = [w_{i1}, w_{i2}, \ldots, w_{in}]^T$ is the weight vector between the i-th hidden neuron and the input neurons, $\beta_i = [\beta_{i1}, \beta_{i2}, \ldots, \beta_{in}]^T$ is the weight vector between the i-th hidden neuron and the outputs, and $b_i$ is the threshold of the *i-th* hidden neuron. The above equation can be written as

$$H\beta = T \tag{3}$$

where

$$H = \begin{bmatrix} g(w_1^T x_1 + b_1) & \cdots & g(w_k^T x_1 + b_k) \\ \vdots & \ddots & \vdots \\ g(w_1^T x_N + b_1) & \cdots & g(w_k^T x_N + b_k) \end{bmatrix}$$

$$\beta = \begin{bmatrix} \beta_1^T \\ \vdots \\ \beta_k^T \end{bmatrix} \text{ and } T = \begin{bmatrix} t_1^T \\ \vdots \\ t_k^T \end{bmatrix}$$

The output weights of ELM are obtained by determining a solution that achieves the least squares error of the linear system. The unique smallest norm least-squares solution can be obtained by

$$\hat{\beta} = H^\dagger T \tag{4}$$

where $H^\dagger$ represents the Moore-Penrose generalized inverse of matrix H. The Moore-Penrose generalized inverse can be calculated by utilizing different approaches such as the orthogonal projection method, iterative method, and singular value decomposition [54]. On the basis of ridge regression theory and the fact that the number of training samples is far bigger than the dimensionality of the feature space, it is better to add a positive value to the diagonal of $H^T T$, which provides a better generalization performance as shown numerically in [55].

$$\hat{\beta} = \left(\frac{I}{C} + H^T H\right)^{-1} H^T T \tag{5}$$

where C is a positive constant. For binary classification, the decision function can be written in the manner shown below

$$f(x) = \text{sign}\left(h(x)\left(\frac{I}{C} + H^T H\right)^{-1} H^T T\right) \tag{6}$$

ELM is used in the second layer of our multi-level system to efficiently detect the Probe attacks.

*3.3. Hierarchical Extreme Learning Machine (H-ELM)*

The Hierarchical Extreme Learning Machine (H-ELM) [56] was proposed in order to achieve a better generalization with faster convergence in comparison with the fundamental ELM approach. H-ELM training is structurally divided into two stages: (i) unsupervised hierarchical feature representation and (ii) supervised feature classification. First, the input instances should be converted into an ELM feature space, which may help in the extraction of hidden information from training instances. Then, high-level sparse features are obtained by applying M-layer unsupervised learning stage. The output of each hidden layer is defined as

$$H_i = g(H_{i-1} \cdot \beta) \tag{7}$$

where $H_i$ is the output of the i-th hidden layer $i \in [1, K]$, $H_{i-1}$ is the output of the (i-1) th hidden layer, g(.) represents the activation function of the hidden layers, and $\beta$ denotes the output weights. After extracting the features of the previous layer, the parameters of the current hidden layer will remain fixed [56]. H-ELM employs random projections of extracted features as the inputs of the feature classification stage. $\ell_1$ penalty is applied to produce sparser and more significant information. The input weights of the ELM sparse autoencoder are obtained by searching the path from a random mapped feature space [56]. The autoencoder enjoys universal approximation capability and sparsity constraint is added to the optimization model, which is written as

$$O_\beta = \underset{\beta}{\text{argmin}}(\|H\beta - X\|^2 + \|\beta\|_{\ell_1}) \tag{8}$$

where X denotes the input data, H represents the random mapping output, and β is the hidden layer weight. A fast iterative shrinkage-thresholding algorithm (FISTA) [57] is used in order to solve Eq. (6). As mentioned before, the supervised training stage is implemented by the original ELM, which is already explained in section 3.2. H-ELM is used in the third, fourth, and fifth layers of our proposed system in order to detect U2R, R2L, and unknown attacks respectively.

**4. Proposed Multi-Level Hybrid IDS**

In this section, we present our proposed system. The system is inspired by the study presented in [52], which uses a five-level classification model based on a combination of SVM and ELM algorithms. As shown in Figure 1, DoS and Probe attacks are detected prior to other categories. This is due to the fact that both DoS and Probe attacks have lower similarity in comparison with other attack types [20]. Conversely, both U2R and R2L attacks are relatively similar to the normal traffic patterns [20] and therefore, both of them present a potential threat that need to be detected before the normal traffic [58].

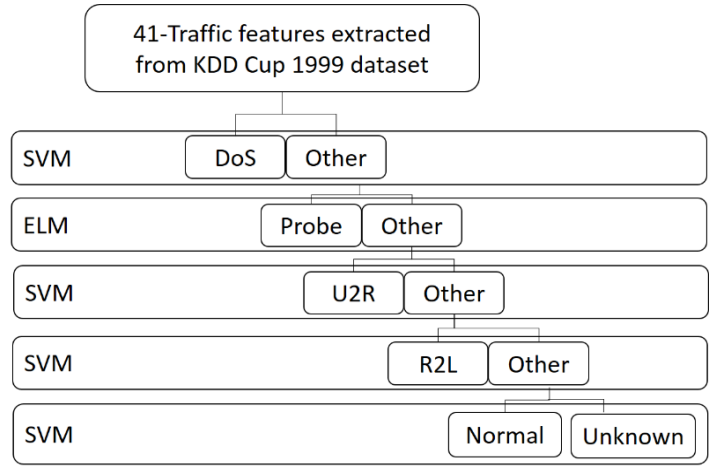

**Figure 1.** Multi-level hybrid IDS proposed in [52]

Similar to the model presented in [52], our proposed system uses one classifier per layer. As shown in Figure 2, the second layer remains the same where ELM achieves better performance as compared to other approaches such as SVM and neural networks [51, 53]. In the next layers, we replace the SVM with H-ELM, which is faster than SVM and performs a better generalization in comparison with SVM and other approaches [51–53]. It is worth noting that in our model we use 6-flow features instead of the 41 traffic-features used in [52], not to mention that our experimental study is conducted on the basis of the NSL-KDD, which is an enhanced version of the KDD Cup 1999.

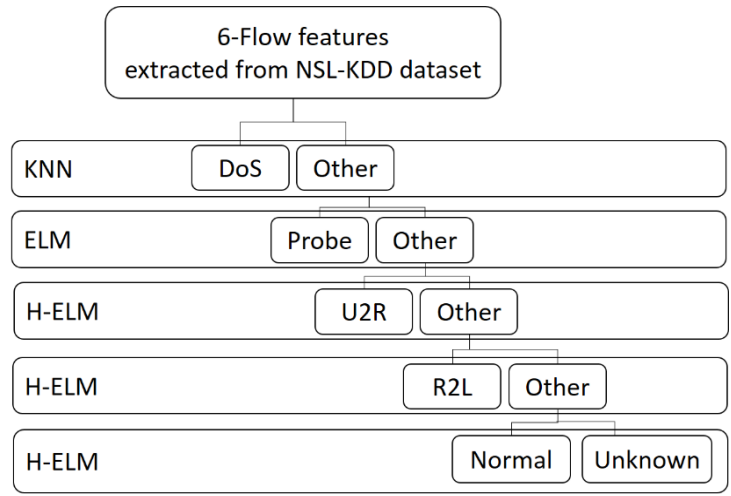

**Figure 2.** The architecture of our proposed IDS

At each layer, the training dataset is spilt into two categories: the first one represents a specific attack type and "Other" category, which represents a normal traffic or an attack that has to be detected by the next layer. Therefore, the model uses five datasets and utilizes the one-versus-all training approach. In the first layer, we use the kNN approach instead of the SVM due to its robustness against noisy training data [34].

## 5. Dataset and Selected Features

As mentioned in the previous sections, in this study, we use the NSL-KDD dataset. The NSL-KDD is an enhanced version of the KDD Cup 99 dataset that suffers from a huge number of redundant records [59]. The NSL-KDD dataset includes the features shown in Table 1. We use features number F1, F2, F5, F6, F23, and F24. The reason for selecting these features is that we can obtain them easily from the SDN controller [32], and therefore can be a good choice for SDNs.

Table 1. List of features of NSL-KDD dataset.

| F# | Feature name | F# | Feature name | F# | Feature name |
|---|---|---|---|---|---|
| F1 | Duration | F15 | Su attempted | F29 | Same srv rate |
| F2 | Protocol type | F16 | Num root | F30 | Diff srv rate |
| F3 | Service | F17 | Num file creations | F31 | Srv diff host rate |
| F4 | Flag | F18 | Num shells | F32 | Dst host count |
| F5 | Source bytes | F19 | Num access files | F33 | Dst host srv count |
| F6 | Destination bytes | F20 | Num outbound cmds | F34 | Dst host same srv rate |
| F7 | Land | F21 | Is host login | F35 | Dst host diff srv rate |
| F8 | Wrong fragment | F22 | Is guest login | F36 | Dst host same src port rate |
| F9 | Urgent | F23 | Count | F37 | Dst host srv diff host rate |
| F10 | Hot | F24 | Srv count | F38 | Dst host serror rate |
| F11 | Number failed logins | F25 | Serror rate | F39 | Dst host srv serror rate |
| F12 | Logged in | F26 | Srv serror rate | F40 | Dst host rerror rate |
| F13 | Num compromised | F27 | Rerror rate | F41 | Dst host srv rerror rate |
| F14 | Root shell | F28 | Srv rerror rate | F42 | Class label |

The NSL-KDD dataset contains a total of 39 attacks wherein each attack is classified into one of the following four categories: DoS, R2L, U2R and Probe. In addition, a set of these attacks is added to the testing set. Table 2 presents the distribution of the known and new attack records in the NSL-KDD testing set.

Table 2. Distributions of known and new attacks in KDD-Test set.

| | DoS | R2L | U2R | Probe |
|---|---|---|---|---|
| Known attacks | 5741 | 2199 | 37 | 1106 |
| | 76.98% | 79.85% | 18.50% | 45.68% |
| New attacks | 1717 | 555 | 163 | 1315 |
| | 23.02% | 20.15% | 81.50% | 54.32% |

## 6. Evaluation Metrics

The performance of our multi-level flow-based system is evaluated in terms of accuracy and False Alarm Rate (FAR), where the accuracy is calculated as

$$Accuracy = \frac{TP + TN}{(TP + TN + FN + FP)} \quad (9)$$

True Positives (TP) is the number of attack instances correctly classified; True Negatives (TN) is the number of normal traffic instances correctly classified; False Positives (FP) is the number of normal traffic instances falsely classified; and False Negatives (FN) is number of attack instances falsely classified. Conversely, false alarm rate is calculated by

$$False\ Alarm\ Rate = \frac{FP}{FP + TN} \quad (10)$$

In addition, we calculate Precision, Recall, and F-measure, which are obtained as

$$Precision = \frac{TP}{TP + FP} \quad (11)$$

$$Recall\ (Detection\ Rate) = \frac{TP}{TP + FN} \quad (12)$$

$$F - measure = 2 \times \frac{(Precision \times Recall)}{(Precision + Recall)} \quad (13)$$

Precision reveals the percentage of attacks detected by an IDS that are actual attacks. Recall indicates the percentage of detected attacks versus all attacks presented in the NSL-KDD dataset. F-measure represents a more balanced measure by considering both precision and recall [32].

## 7. Experimental Results

The experiment study was conducted on an Intel i5 machine with 12 GB of RAM. Table 3 shows the parameters used for each classifier employed at each layer.

Table 3. Parameters used for each classifier at each layer

| Layer | Classifier | Parameter (s) | | |
|---|---|---|---|---|
| 1 | kNN | K = 65 | | |
| 2 | ELM | N = 400 | | |
| 3 | H-ELM | N1 = 40 | N2 = 40 | N3 = 200 |
| 4 | H-ELM | N1 = 10 | N2 = 10 | N3 = 300 |
| 5 | H-ELM | N1 = 10 | N2 = 10 | N3 = 200 |

We divided our experimental study into two experiments: 1) evaluation of the performance at each layer and 2) comparison with other supervised machine learning approaches. The first experiment basically evaluated the performance of the model. Table 4 shows the accuracy and the false alarm rate achieved at each layer.

Table 4. Results for training and testing stages achieved at each layer of our proposed system

| Layer | Classifier | Detected Attack | Training Stage | | Testing Stage | |
|---|---|---|---|---|---|---|
| | | | Accuracy (%) | False Alarm Rate (%) | Accuracy (%) | False Alarm Rate (%) |
| 1 | kNN | DoS | 97.34 | 1.38 | 91.23 | 3.39 |
| 2 | ELM | Probe | 97.12 | 0.12 | 92.61 | 1.45 |
| 3 | H-ELM | U2R | 99.96 | 0.0024 | 99 | 0.12 |
| 4 | H-ELM | R2L | 99.19 | 0.0240 | 86.97 | 0.94 |
| 5 | H-ELM | Unknown | 90.76 | 7.28 | 80.39 | 5.61 |
| All layers | kNN + ELM + H-ELM | All types | 94.11 | 7.89 | 84.29 | 6.3 |

In the second experiment, we compared our proposed system with other conventional supervised learning approaches such as Naive Bayes, Neural Networks, SVM, and Decision Trees. As displayed in Table 5, our system achieves the highest values of accuracy, recall, and F1-score. In terms of false positive rate, it is worth mentioning that the H-ELM approach achieves the highest precision with the lowest false alarm rate. Compared with [32], our system achieved a higher accuracy, precision, recall, and F1-score. However, in terms of false alarm rate, [32] achieved better results.

Table 5. Comparing our proposed approach with other approaches

| Method | Accuracy (%) | False Alarm Rate (%) | Precision (%) | Recall (%) | F1-score (%) |
|---|---|---|---|---|---|
| Naive Bayes | 49.88 | 5.14 | 80.28 | 15.83 | 26.45 |
| Neural Network | 63.70 | 7.66 | 87.88 | 42.02 | 56.86 |
| SVM | 71.40 | 10.63 | 87.79 | 57.80 | 69.71 |
| Decision Trees | 74.43 | 6.43 | 92.50 | 59.95 | 72.75 |
| ELM (N=1500) | 74.80 | 10.17 | 89.18 | 63.43 | 74.13 |
| Deep Neural Network [32] | 75.75 | 3.21 | 83 | 76 | 75 |
| kNN | 77.09 | 4.07 | 95.33 | 62.84 | 75.75 |
| H-ELM (N1=30,N2=30,N3=300) | 77.59 | **2.57** | **96.98** | 62.59 | 76.08 |
| H-ELM (N1=10,N2=10,N3=200) | 80.39 | 5.61 | 94.26 | 69.80 | 80.21 |
| **Our Approach** | **84.29** | 6.3 | 94.18 | **77.18** | **84.83** |

We also used the Cbench tool in order to evaluate the throughput of our system. Therefore, we implemented the system as a module of POX controller in the SDN's control plane. This approach is considered more efficient than the implementation of the system as an application of the controller due to the fact that our 6-flow features can be easily obtained by the controller, and increasing number of flows will dramatically increase the interaction between the controller and the application. Previously mentioned features were collected periodically every 10 seconds. Each Open vSwich was connected to 1000 virtual hosts with different MAC addresses, sending 10000 (Packet-In) messages to the POX controller. Compared with the basic forwarding module, our system achieved an acceptable performance as presented in Figure 3. This is owing to the fact that we use only 6-flow features that can easily be obtained from the controller and employ machine learning algorithms that reduce the time required for the testing stage such as ELM and H-ELM in the layers from 2 to 5.

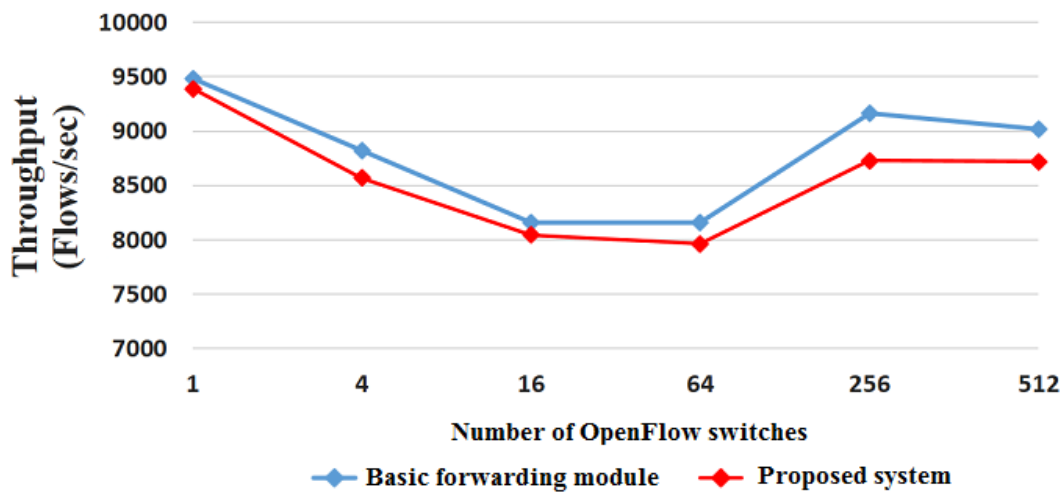

**Figure 3.** The throughput of our proposed IDS and the basic forwarding module

In brief, our experimental study showed that both common supervised and deep neural network [32] approaches may not always achieve the best results, and that the usage of alternative methods such as multi-level systems could be investigated in order to attain the required results. Second, achieving good results through a multi-level approach does not guarantee good results in terms of detecting unknown attacks such as the results obtained from [52]. Third, achieving a high accuracy through a multi-level system may be accompanied by an increased number of false alarms, which is observed in our proposed system.

## 8. Conclusions

In this paper, we proposed an efficient multi-level hybrid intrusion detection method for SDNs. The system was designed on the basis of a combination of kNN, ELM, and H-ELM approaches. The experimental study conducted on NSL-KDD dataset showed that our approach significantly enhanced the overall accuracy. Our future work will be focused on improving the system in order to achieve a lower false alarm rate.